\documentstyle[12pt,twoside]{article}
{\catcode `\@=11 \global\let\AddToReset=\@addtoreset}
\AddToReset{equation}{section}

\def\greaterthansquiggle{\raise.3ex\hbox{$>$\kern-.75em\lower1ex\hbox{$\sim$}}}
\def\lessthansquiggle{\raise.3ex\hbox{$<$\kern-.75em\lower1ex\hbox{$\sim$}}}
\newcommand{\beq}{\begin{equation}}
\newcommand{\eeq}{\end{equation}}
\newcommand{\beqa}{\begin{eqnarray}}
\newcommand{\eeqa}{\end{eqnarray}}
\newcommand{\beqan}{\begin{eqnarray*}}
\newcommand{\eeqan}{\end{eqnarray*}}
\newcommand{\ba}{\begin{array}}
\newcommand{\ea}{\end{array}}

\newcommand{\ol}{\overline}
\newcommand{\ra}{\rightarrow}

\newcommand{\ve}{\varepsilon}
\newcommand{\vp}{\varphi}

\newcommand{\dg}{\dagger}
\newcommand{\wt}{\widetilde}
\newcommand{\wh}{\widehat}

\newcommand{\A}{{\cal A}}

\newcommand{\F}{{\cal F}}

\newcommand{\Ha}{{\cal H}}

\newcommand{\st}{\stackrel}

\def\nz{\ifmmode {I\hskip -3pt N} \else {\hbox {$I\hskip -3pt N$}}\fi}
\def\zz{\ifmmode {Z\hskip -4.8pt Z} \else
       {\hbox {$Z\hskip -4.8pt Z$}}\fi}
\def\qz{\ifmmode {Q\hskip -5.0pt\vrule height6.0pt depth 0pt
       \hskip 6pt} \else {\hbox
       {$Q\hskip -5.0pt\vrule height6.0pt depth 0pt\hskip 6pt$}}\fi}
\def\rz{\ifmmode {I\hskip -3pt R} \else {\hbox {$I\hskip -3pt R$}}\fi}
\def\cz{\ifmmode {C\hskip -4.8pt\vrule height5.8pt\hskip 6.3pt} \else
       {\hbox {$C\hskip -4.8pt\vrule height5.8pt\hskip 6.3pt$}}\fi}

\normalsize
\begin{document}
\bibliographystyle{plain}
\begin{titlepage}
\begin{flushright}
UWThPh-1997-25\\
\today
\end{flushright}
\vspace{2cm}
\begin{center}
{\Large \bf Refined Algebraic Quantization and \\[5pt]
Quantum Field Theory in Curved Space-Time  }\\[50pt]
Helmut Rumpf  \\
Institut f\"ur Theoretische Physik \\ Universit\"at Wien\\
Boltzmanngasse 5, A-1090 Wien, Austria
\vfill
{\bf Abstract} \\
\end{center}

Application of the so-called refined algebraic quantization scheme 
for constrained systems to the relativistic particle provides an inner
product that defines a unique Fock representation for a scalar field 
in curved space-time. The construction can be made rigorous for a general
globally hyperbolic space-time, but the quasifree state so obtained
turns out to be unphysical in general. We exhibit a closely related
pair of Fock representations that is also defined generically and
conforms to the notion of in- and outgoing states in those situations
where particle creation by the external field is expected.

\vfill 
\end{titlepage}

\section{Introduction}
In the early years of quantum field theory in curved space-time the two
most important foundational problems were deemed to be the following:
First, how to generalize the notion of vacuum in Minkowski space to
space-times with a lesser degree of symmetries, and second, how to get
rid of the divergencies that appear in the expectation value of the
energy-momentum tensor and related objects of physical interest. After
more than thirty years of development, which was most rapid in the years
immediately following the discovery of the Hawking effect in 1974, a
certain stage of maturity has been reached, and the following consensus
regarding the above problems appears to have emerged. First, particle-like
states may not exist in a generic space-time, and quantum field theory 
should be formulated in a manner that is not tied to a particular Fock
representation, just as General Relativity does not require the use of a
particular coordinate system. Second, a consistent regularization scheme
exists for a large class of states called Hadamard states, and it has been
proposed that there are no physical states outside this class.

In this paper, without questioning the answer to the first point raised
above, we nonetheless would like to draw attention to the fact that
there exist mathematically preferred states which are invariantly
defined and have physical significance. This is completely analogous
to the existence of preferred reference frames in General Relativity, 
e.g. free falling ones, that are more ``physical'' than others. We shall
infer the existence of preferred states (which have been known for quite
a long time) from the so-called refined algebraic quantization scheme
of Ashtekar et al. \cite{AL} which has been devised in the context of the
connection dynamics formulation of canonical quantum gravity, but will be
applied here only to a very simple system, namely the relativistic particle.
Although this application falls within relativistic quantum mechanics,
not quantum field theory, it is straightforward to associate a unique
Fock representation with it. This turns out to be unphysical in general,
however. But a closely related pair of Fock representations appears to
correspond exactly to the notions of in- and outgoing states in those
situations where particle creation by the external field is expected.
Moreover it is a characteristic of these states that they allow a simple
description of particle creation within the framework of relativistic
quantum mechanics after all.

It is our aim to present a mathematically rigorous construction of the
various representations. Therefore some necessary preliminaries of
quantum field theory in curved space-time are recapitulated in Sec.~2,
and the nuclear spectral theorem plays a prominent role in the
application of the refined algebraic quantization method given in Sec.~3.
Physical considerations enter more directly in Sec.~4, but for the
application of the formalism to concrete physical situations we have to
refer to the published literature.

\section{Linear scalar field in curved space-time}
We begin with a review of the standard construction of quantum field
theory in curved space-time as formulated by Kay and Wald \cite{KW,W}.
Let $(M,g_{ab})$ be a globally hyperbolic space-time manifold. This
implies that $M$ is a foliation of spacelike Cauchy hypersurfaces,
\beq
M = \bigcup_t \Sigma_t,
\eeq
when $t$ is a global time coordinate. We shall consider a scalar field
$\phi$ on $M$ whose classical action is of the form
\beqa
S[\phi] &=& \frac{1}{2} \int d^4x |g|^{1/2} (g^{ab} \nabla_a \phi
\nabla_b \phi - V(x) \phi^2) \\
&\equiv& \int dt L[\phi,\dot \phi]
\eeqa
where the dot denotes differentiation with respect to $t$. Variation of
the action gives rise to the classical field equation
\beqa
\wh C \; \phi &=& 0 \\
\wh C &:=& \Box_g + V \\
\Box_g &=& g^{ab} \nabla_a \nabla_b = |g|^{-1/2} \partial_a |g|^{1/2}
g^{ab} \partial_b.
\eeqa
A popular choice for the ``potential'' $V$ is
\beq
V = \frac{1}{6} R + m^2
\eeq
implying conformal invariance for the massless field at the classical 
level. The Hamiltonian formalism for the field $\phi$ becomes most
transparent upon a $3+1$ decomposition of the metric. In particular one
introduces the induced Riemannian 3-metric on $\Sigma_t$, denoted by
$h_{ij}$, and the future-directed unit normal vector field $n^a$ on
$\Sigma_t$. Then the canonically conjugate momentum of $\phi$ may be
expressed as
\beq
\Pi = \frac{\delta L}{\delta \dot \phi} = |h|^{1/2} n^a \nabla_a \phi ,
\eeq
$h$ denoting the determinant of $h_{ij}$. We define the phase space
$\Gamma$ as
\beq
\Gamma = \{ (\vp,\pi)|\vp,\pi \in C_0^\infty(\Sigma_0)\}.
\eeq
Because of global hyperbolicity we have
\beq
\Gamma \cong S
\eeq
where $S$ is the space of classical solutions of (1.4) with $C_0^\infty$
initial data. Moreover the linearity of the field equation implies
$\Gamma \cong T_P \Gamma$ (the tangent space at an arbitrary point
$P \in \Gamma$) so that the canonical 2-form defines a natural symplectic
form $\Omega : S \times S \ra {\bf R}$. It is given by
\beq
\Omega(\vp_1,\vp_2) = \int_{\Sigma_t} (\vp_1 \pi_2 - \pi_1 \vp_2)d^3x
= \int_{\Sigma_t} \vp_1 \st{\leftrightarrow}{\nabla}_a \vp_2
d \sigma^a
\eeq
and clearly independent of $t$. (The hypersurface element is
$d \sigma^a = n^a |h|^{1/2} d^3x$ in local coordinates on $\Sigma_t$.)
Classical observables are functionals on $S$, i.e. maps from $S$ to
{\bf R}. The physically most prominent one, the point field
\beq
\Phi(x) : \vp \mapsto \vp(x)
\eeq
is a distribution, which becomes an operator-valued distribution upon
quantization and is therefore not considered for constructive purposes.
Instead, one considers the ``symplectically smeared'' fields
\beq
\Omega_\psi \equiv \Omega(\psi,\cdot) : \vp \mapsto \Omega(\psi,\vp).
\eeq
(If one lifts the restriction of the solution space to $C_0^\infty$
initial data, then one has the following relations between the smeared
and point fields: $\Phi(x) = \Omega_{G(x,\cdot)}$; $\Omega_\psi =
\Omega(\psi,\Phi(\cdot))$, where $G$ denotes the fundamental solution
or classical propagator.) We define the classical algebra of observables
as the commutative algebra generated by the smeared fields. Its Poisson
structure is implied by the canonical Poisson bracket:
\beq
\{ \Omega_{\psi_1},\Omega_{\psi_2}\} = \Omega(\psi_1,\psi_2).
\eeq

Canonical quantization introduces a quantum algebra of observables with
generators $\wh \Omega_\psi$ and the canonical commutation relations
\beq
[ \wh \Omega_{\psi_1},\wh \Omega_{\psi_2}] = i \Omega(\psi_1,\psi_2)
{\bf 1}
\eeq
(we set $\hbar = 1$). States are positive normed linear functionals on this
algebra. The physically most interesting states are defined in terms of
Fock space constructions. A Fock space may be constructed in the
following way: Select a space $S^{{\bf C}_+}$ of complex solutions of
(2.4) ($S^{{\bf C}_+} \cap \ol{S^{{\bf C}_+}} = \{0\}$) such that (i) there 
is a linear bijective map $P^+ : S \ra S^{{\bf C}_+}$ with
$\vp = P^+ \vp + \ol{P^+ \vp}$ $\forall \; \vp \in S$, (ii) $\Omega$
is extendible to $S^{{\bf C}_+} \oplus \ol{S^{{\bf C}_+}}$ and (iii)
the charge form, defined by
\beq
(\psi_1,\psi_2) := i \Omega(\ol{\psi}_1,\psi_2)
\eeq
is positive on $S^{{\bf C}_+}$ and $(S^{{\bf C}_+},\ol{S^{{\bf C}_+}})
= 0$. (Note that $S^{{\bf C}_+}$ is not required to be a subspace of the
complexification of $S$, ${\bf C} \otimes S$, the reason being that this 
would not yield a physical one-particle Hilbert space even in Minkowski
space-time.) The one-particle Hilbert space $\Ha^+$ is defined as the
completion of $S^{{\bf C}_+}$ with respect to the charge form. It follows
from (iii) that
\beqa
\left.(\;,\;)\right|_{\Ha^+} &\geq& 0 \\
\left.(\;,\;)\right|_{\ol{\Ha^+}} &\leq& 0 \\
(\Ha^+, \ol{\Ha^+}) &=& 0.
\eeqa
(Eq. (2.18) is implied by $(\ol{\vp},\ol{\psi}) = - (\psi,\vp)$.)
From $\Ha^+$ we construct a representation $\rho$ of the algebra of
observables in the Fock space
\beq 
\F_s(\Ha^+) = \bigoplus_{n=0}^\infty \otimes{}_s^n \; \Ha^+
\eeq
with
\beq
\rho(\wh \Omega_\vp) = - ia(\ol{P^+\vp}) + i a^\dg(P^+ \vp)
\eeq
where $a(\psi)$ and $a^\dg(\psi)$ are the standard annihilation and
creation operators associated with $\psi \in \Ha^+$ with commutator
\beq
[a(\psi_1),a^\dg(\psi_2)] = (\psi_1,\psi_2) {\bf 1}.
\eeq
Obviously the whole construction depends on the choice of $S^{{\bf C}_+}$;
different choices may yield unitarily inequivalent representations.

It turns out that any choice of $S^{{\bf C}_+}$ is characterized by a
certain positive definite inner product on $S$, and that this
characterization provides a very useful reformulation of the above
construction. In order to prepare this reformulation we observe that a
positive inner product $\langle \;,\; \rangle$ on $S$ may be defined by
\beq
\langle \vp_1,\vp_2\rangle := 2 \mbox{Re }(P^+ \vp_1,P^+ \vp_2)
\eeq
so that
\beq
(P^+\vp_1,P^+\vp_2) = \frac{1}{2} \; \langle \vp_1,\vp_2 \rangle + 
\frac{i}{2}\; \Omega(\vp_1,\vp_2)
\eeq
(the imaginary part of the expression appearing in (2.24) is independent
of the choice of $S^{{\bf C}_+}$). Using the Schwarz inequality for the
charge form on $S^{{\bf C}_+}$,
\beq
(\psi_1,\psi_1)^{1/2} (\psi_2,\psi_2)^{1/2} \geq |(\psi_1,\psi_2)|
\geq |\mbox{Im}(\psi_1,\psi_2)|,
\eeq
for $\psi_i = P^+ \vp_i$ and noting that this chain of inequalities can be
saturated, we deduce
\beq
\langle \vp_1,\vp_1\rangle = \sup_{\vp_2 \neq 0} 
\frac{[\Omega(\vp_1,\vp_2)]^2}{\langle \vp_2,\vp_2\rangle}.
\eeq

The Fock construction may now be reformulated as follows. Let
$\langle \; , \;\rangle$ be a positive definite scalar product on $S$ with
the ``supremum property'' (2.26) relative to the charge form. Complete
$S$ with respect to $\langle \;,\;\rangle$ to obtain a real Hilbert space
$\wh S$. Eq. (2.26) implies that $\Omega : S \times S \ra {\bf R}$ is
bounded and by continuity extendible to $\wh S \times \wh S$. Define
$J : \wh S \ra \wh S$ by
\beq
\Omega(\vp_1,\vp_2) = \langle \vp_1,J \vp_2 \rangle .
\eeq
One verifies easily that $J^\dg = -J$, $J^2 = - {\bf 1}$ (i.e. $J$ defines a
complex structure, but it will not be used in this sense). Next complexify
$\wh S$ to $\wh S^{\bf C} = {\bf C} \otimes \wh S$ and extend $\Omega$
(bilinear), $\langle\;,\;\rangle$ (sesquilinear) and $J$ to $\wh S^{\bf C}$ by
complex linearity. Define the complex Hilbert space $\Ha^+$ as the
eigenspace of $J$ with
\beq
\left. J\right|_{\Ha^+} = - i
\eeq
and let $\wt P^+$ be the orthogonal projection from $\wh S^{\bf C}$ on $\Ha^+$.
It can then be verified that $\Ha^+$ has the properties of a one-particle
Hilbert space. Moreover defining
\beq
P^+ = \left. \wt P^+\right|_S
\eeq
we recover (2.24). Thus the choice of a complex solution space 
$S^{{\bf C}_+}$ is indeed equivalent to the choice of a certain inner
product $\langle \;,\;\rangle$ on $S$ with the property (2.26).

The above construction may be generalized to represent general
quasifree states (of which the Fock vacua form only a subclass).
These states are best introduced on the Weyl algebra $\A$, a subalgebra
of the algebra of observables that is generated by the unitary elements
\beq
\wh W_\vp = e^{i \wh \Omega_\vp}.
\eeq
The canonical commutation relations (2.15) imply the Weyl relations
\beq
\wh W_{\vp_1} \wh W_{\vp_2} = e^{\frac{i}{2} \Omega(\vp_1,\vp_2)}
\wh W_{\vp_1 + \vp_2}.
\eeq
Let $\langle \;,\;\rangle : S \times S \ra {\bf R}$ be positive with
\beq
\langle \vp_1,\vp_1\rangle \; \langle \vp_2,\vp_2 \rangle \geq
[\Omega(\vp_1,\vp_2)]^2.
\eeq
This inequality is a certain relaxation of the supremum condition (2.26)
and will therefore yield a wider class of Fock space constructions than
considered previously. The inner product $\langle \; ,\;\rangle$ defines
a quasifree state $\omega$ by
\beq
\omega(\wh W_\vp) = e^{- \frac{1}{2}\;\langle \vp,\vp\rangle}.
\eeq
Any quasifree state (meaning a state whose truncated $n$-point functions
vanish for $n > 2$) may be related to an inner product in this way, the
inequality (2.32) ensuring the positivity of the state functional.
Although $\wh \Omega_\vp$ does not belong to the Weyl algebra, the
``smeared two-point function'' may be defined by
\beq
\langle \wh \Omega_{\vp_1} \wh \Omega_{\vp_2}\rangle_\omega =
- \frac{\partial^2}{\partial s \partial t}
\left.\left[ \omega(W_{s\psi_1 + t \psi_2})
e^{ist \Omega(\psi_1,\psi_2)/2} \right] \right|_{s=t=0}
\eeq
and eq. (2.33) implies
\beq
\langle \wh \Omega_{\vp_1} \wh \Omega_{\vp_2}\rangle_\omega =
\frac{1}{2} \; \langle \vp_1,\vp_2 \rangle + 
\frac{i}{2} \; \Omega(\vp_1,\vp_2).
\eeq
The GNS construction for $\omega$ has a natural Fock space structure, i.e.
\beq
\omega(\wh A) = \langle 0|\rho(\wh A)|0\rangle \qquad \forall \;
\wh A \in \A
\eeq
where 
\beq
\rho(\wh W_\vp) = \exp [a(\ol{P^+\vp}) - a^\dg(P^+\vp)]
\eeq
and the Fock vacuum $|0\rangle$ is the cyclic vector of the representation.
But in general $\Ha^+$ is not simply a space of complex solutions, and
only $P^+S + iP^+ S$ (and not $P^+S$ itself as in the case of a pure state
$\omega$) is dense in $\Ha^+$. The following four statements can be proven
to be equivalent (see \cite{KW}): (i) $\omega$ is pure, (ii) $\rho$ is
irreducible, (iii) $P^+S$ is dense in $\Ha^+$, (iv) the inner product
$\langle \;,\;\rangle$ obeys the supremum condition (2.26).

\section{A ``physical'' Hilbert space of classical solutions}
In this section a certain candidate for an inner product obeying (2.32)
or even (2.26) will be constructed. We consider this candidate to be a
natural one because it is provided by a certain refinement of the Dirac
quantization prescription for constrained systems. This refinement was
proposed under the name ``refined algebraic quantization'' by Ashtekar
et al. \cite{AL} in the context of the connection dynamics approach to
canonical quantum gravity. It is in fact a physicist's version of what
is known as Rieffel induction in pure mathematics \cite{L}. The latter 
may be considered to be a quantum analog of Marsden--Weinstein reduction
(a geometrical construction of a reduced phase space) and of a more
general method of constructing new symplectic spaces and Poisson morphisms 
from old ones \cite{X}.

Rather than restating the general principles of the refined algebraic
quantization scheme (which can be found in \cite{AL}) we shall confine
ourselves here to a simplified version that is sufficient for its
application to the relativistic particle, and its main points should
become clear from the application itself. We start from the
reparametrization-invariant action for the relativistic particle in a
curved space-time, admitting also a nontrivial ``potential'' $V$:
\beq
S[x(\tau)] = - \int [V(x)]^{1/2} ds.
\eeq
This contains the invariant line element
\beq
ds = \left( g_{ab} \frac{dx^a}{d\tau} \frac{dx^b}{d\tau}\right)^{1/2}
d\tau
\eeq
and implies the constraint
\beq
C \equiv - g^{ab} p_a p_b + V(x) = 0.
\eeq
Quantization according to the refined algebraic scheme proceeds in two
steps.

Step 1 consists in the quantization of the ``unconstrained'' system. In
the position representation state vectors are represented by wave functions
$\wt \psi \in L^2(M,d^4x)$ and the canonically conjugate momenta by 
operators
\beq
\wh p_a : \wt \psi \ra i \frac{\partial}{\partial x^a} \; \wt \psi.
\eeq
Note that it is the requirement that these operators be self-adjoint
that distinguishes the non-invariant measure $d^4x$ defining the $L^2$
space. As scalar products have to be invariant, $|\wt \psi|^2$ must be a
scalar density. Equivalently, however, we can define scalar wave functions
\beq
\psi := |g|^{-1/4} \wt \psi
\eeq
which are acted upon by the conjugate momenta as
\beq
\wh p_a : \psi \ra |g|^{-1/4} i \frac{\partial}{\partial x^a} |g|^{1/4}
\psi.
\eeq
These wave functions $\psi$ are elements of an invariantly defined Hilbert
space, which has only auxiliary status, however, and will therefore be
denoted by $H_{\rm aux}$:
\beq
\psi \in H_{\rm aux} \equiv L^2(M,|g|^{1/2} d^4x).
\eeq
The relevant inner product is given by
\beq
\langle \vp,\psi\rangle_{\rm aux} = \int d^4x \; |g|^{1/2} \ol{\vp} \psi.
\eeq
The quantum version of the constraint requires the factor ordering
\beq
(g^{ab} p_a p_b)\wh{\;} = |g|^{-1/4} \wh p_a |g|^{1/2} g^{ab} \wh p_b
|g|^{-1/4}
\eeq
so as to yield the local differential operator $\wh C$ of (2.5) in the
position representation. The operator $\wh C$ is unbounded in $H_{\rm aux}$
and in general the Dirac quantization condition
\beq
\wh C\;\psi = 0
\eeq
has no solution in $H_{\rm aux}$.

Step 2 of the refined algebraic quantization scheme defines formally a 
``physical'' Hilbert space of solutions of (3.10). It starts from the
subspace
\beq
D_M \equiv C_0^\infty (M) \subset H_{\rm aux}
\eeq
which is invariant under $\wh C$ provided that $V(x)$ is sufficiently
well-behaved. $D_M$ is a nuclear space, the definition of its topology
being a straightforward generalization from the case when $M$ is flat
(note that there is no such straightforward generalization of Schwartz
space to a curved manifold). Since the nuclear topology of $D_M$ is
stronger than that induced by the $L^2$ norm, the nuclear dual $D'_M$
is larger than $H_{\rm aux}$. As the inner product 
$\langle \;,\;\rangle_{\rm aux}$ is continuous with respect to the
nuclear topology, there exists an antilinear embedding of $D_M$ in
$D'_M$, $\vp_0 \mapsto \vp'_0$, defined by
\beq
\vp'_0(\psi_0) := \langle \vp_0,\psi_0\rangle_{\rm aux}.
\eeq
A further antilinear map $\eta : D_M \ra D'_M$ called ``rigging map'',
is defined formally by
\beq
\eta : \psi_0 \mapsto \psi = 2 \pi \delta(\wh C) \ol{\psi}_0
\eeq
whence
\beq
\wh C \;\psi = 0.
\eeq
A formally positive definite inner product $\langle \;,\;\rangle_{\rm phys}$
on $\eta(D_M)$ can be defined by
\beq
\langle \vp,\psi\rangle_{\rm phys} := (\eta \psi_0)(\vp_0).
\eeq
Finally a Hilbert space $H_{\rm phys}$ is obtained by completing
$\eta(D_M)$ with respect to $\langle \;,\;\rangle_{\rm phys}$. This
completes the refined algebraic quantization of the relativistic particle.
Before discussing its physical significance, we turn to the construction 
of the formal definitions involved in step~2.

The construction rests on the assumption that the operator $\wh C$ is
self-adjoint on a dense domain $D \subseteq H_{\rm aux}$ with $D_M
\subseteq D$. (Since $\wh C$ is a real operator, the existence of
self-adjoint extensions is ensured by a theorem of von Neumann.) Thus
$\wh C$ is continuous on $D_M$ w.r.t. the nuclear topology. As
$\wh C D_M \subseteq D_M$, one may define the operator $\wh C' : D'_M
\ra D'_M$ by
\beq
(\wh C' f)(\vp) = f(\wh C\vp).
\eeq
The assumption just stated together with the fact that 
$\langle \;,\; \rangle_{\rm aux}$ is continuous on $D_M$ and $D_M$ is
dense in $H_{\rm aux}$ allows the application of the nuclear spectral
theorem (\cite{GV,M}) with the following result:

There exists a system $\{e_{\lambda k}\}$ of eigenfunctionals of
$\wh C$ (i.e. $\wh C' e_{\lambda k} = \lambda e_{\lambda k}$) such that
for any $\vp_0 \in D_M$ $\vp'_0$ may be represented as
\beq
\vp'_0 = \int_\sigma \sum_{k=1}^{m_\lambda} \ol{\wt \vp}_{\lambda k}
e_{\lambda k} d\mu(\lambda)
\eeq
where $\sigma$ is the spectrum of $\wh C$, $\mu$ is a measure and
$m_\lambda$ the multiplicity of the spectral value $\lambda$ (as they
are defined in the multiplication operator version of the spectral
theorem). Moreover
\beq
\wt \vp_{\lambda k} = e_{\lambda k}(\vp_0).
\eeq

From (3.17) we infer the exact definition of the rigging map (3.13):
\beq
\vp \equiv \eta \vp_0 = 2 \pi \sum_{k=1}^{m_0} \wt \vp_{0k} e_{0k}.
\eeq
Hence the inner product (3.15) is expressed as
\beq
\langle \vp,\psi\rangle_{\rm phys} = (\eta \psi_0)(\vp_0) =
2 \pi \sum_{k=1}^{m_0} \wt \vp_{0k} \ol{\wt \psi}_{0k}
\eeq
where we have used (3.18).

If the spectrum of $\wh C$ has no singular continuous part, an equivalent
definition of $H_{\rm phys}$ is implied by the fact that in the spectral
representation the $e_{\lambda k}$ are distributions concentrated in a
point $\lambda \in \sigma$. We define the spectral $\delta$ distribution
$\delta_\mu(\lambda,\lambda')$ by
\beq
\int_\sigma d\mu(\lambda') \delta_\mu(\lambda,\lambda') \wt \vp(\lambda')
= \wt \vp(\lambda)
\eeq
for any test function $\wt \vp$. If $\lambda$ belongs to the pure point
spectrum,
\beq
\delta_\mu(\lambda,\lambda') = \delta_{\lambda,\lambda'}.
\eeq
If $\lambda$ is in the absolutely continuous (w.r.t. Lebesgue measure)
part of the spectrum and hence $d\mu/d\lambda \neq 0$ exists, then
\beq
\delta_\mu(\lambda,\lambda') = \left( \frac{d\mu}{d\lambda}\right)^{-1}
\delta(\lambda - \lambda').
\eeq
Note that $\delta(\lambda,\lambda')$ is not translation invariant in
general.

The inner product $\langle e_{\lambda k},e_{\lambda' k'}\rangle_{\rm aux}$
is defined in the above distributional sense. We may choose the system
$\{e_{\lambda k}\}$ to be orthonormal:
\beq
\langle e_{\lambda k},e_{\lambda' k'}\rangle_{\rm aux} =
\delta_\mu(\lambda,\lambda') \delta_{k,k'}.
\eeq
It turns out in all known applications (although apparently there is no
general proof) that the $e_{\lambda k}$ may be chosen such that they are
represented by locally integrable eigenfunctions $e_{\lambda k}(x)$
according to
\beq
e_{\lambda k}(\vp) = \int d^4x |g|^{1/2} e_{\lambda k}(x) \vp(x).
\eeq
Hence the inner product
\beq
\int d^4x |g|^{1/2} \ol{e}_{\lambda k} e_{\lambda' k'} =
\delta_\mu(\lambda,\lambda') \delta_{k,k'}
\eeq
exists as a spectral distribution. Let now $\vp^{(\lambda)}$ be a 
locally integrable solution of $\wh C \vp^{(\lambda)} = \lambda
\vp^{(\lambda)}$. It follows from (3.17), (3.20), (3.25) and (3.26)
that
\beq
\int d^4x |g|^{1/2} \vp^{(0)} \ol{\psi}^{(\lambda)} = 2\pi
\delta_\mu(0,\lambda)\; \langle \vp^{(0)},\psi^{(0)}\rangle_{\rm phys}.
\eeq
Thus the inner product $\langle \;,\;\rangle_{\rm phys}$ turns out to be
the general version of an inner product first proposed by Nachtmann in
the special case of the Klein--Gordon equation on de Sitter space
\cite{N}. The definition (3.27) of the physical inner product is more
useful for practical computation although it requires to embed a given
solution in a whole family parametrized by $\lambda$ (usually, when
$V$ is of the form (2.7) and $m^2 \neq 0$, any exact solution of (2.4)
obtained by analytical methods will appear as an analytical function of
$\lambda$).

As to the physical significance of $H_{\rm phys}$, we remark that in
general the complexified solution space of Sec.~2, ${\bf C} \otimes S$,
is a subspace and the charge form $(\;,\;)$ is defined on $H_{\rm phys}$.
A notable exception are the exponentially growing solutions 
(``resonances'') that may exist in a certain type of external electrostatic
potential in flat space-time \cite{R2} and that may lie in the (then
complex) space $S$, but not in $H_{\rm phys}$. A gravitational analog of 
this situation occurs in de Sitter space \cite{R3}. This suggests an
invariant definition of ``exponentially growing'' or ``complex frequency''
solutions $\psi$ of (2.4): They do not belong to $H_{\rm phys}$, or
equivalently, $\int d^4x |g|^{1/2} \ol{\psi}_0 \psi_\lambda$ does not
exist as a distribution in $\lambda$. The Hilbert space $H_{\rm phys}$
is somewhat reminiscent of the notion of tempered distributions, but
note that $H_{\rm phys}$ is not a nuclear space and that its closure in
$D'_M$ is larger (the closure of the linear span of the system
$\{ e_{\lambda k}\}$ is larger still, namely identical to $D'_M$ itself).

\section{Invariantly defined states}
The inner product $\langle \;,\;\rangle_{\rm phys}$ constructed in the
previous section may be substituted for the inner product
$\langle \;,\;\rangle$ appearing in the Fock space construction of
Sec.~2, and we may investigate the physical meaning of the Fock
space defined invariantly in this way. First we have to check the
positivity condition (2.32) for the states so defined. To this end we
define an operator $N$ on $H_{\rm phys}$ by
\beq
\langle \psi_1,N \psi_2\rangle_{\rm phys} = (\psi_1,\psi_2).
\eeq
It corresponds to the operator $iJ$ of Sec. 2 and is formally self-adjoint.
If (2.32) holds, then
\beq
\frac{|\langle \psi_1,N \psi_2\rangle_{\rm phys}|}{\|\psi_1\|\|\psi_2\|}
\leq 1
\eeq
and hence
\beq
|N| \leq 1.
\eeq
This inequality cannot be proven in full generality, but it appears to be
a generic property. This will be seen from the integral representation of
$N$ that we are going to construct now. As is well known, the $\delta$
distribution appears in the boundary value of a holomorphic function, viz.
\beq
\frac{1}{x+i0} \equiv \lim_{\ve \downarrow 0} \frac{1}{x+i\ve} =
P\; \frac{1}{x} + i \pi \delta(x)
\eeq
($P$ denoting the principal value). Therefore we obtain for the
self-adjoint operator $\wh C$ the relation
\beq
2\pi i \delta(\wh C) = (\wh C - i0)^{-1} - (\wh C + i0)^{-1}
\eeq
(note that $(\wh C \pm i\ve)^{-1}$ is bounded and holomorphic in $\ve$
for $\ve > 0$).

We define the Feynman propagator $K(x,x')$ as the (singular) integral
kernel of $(\wh C - i0)^{-1}$ with respect to space-time integration:
\beq
[(\wh C - i0)^{-1} \psi](x) = - \int d^4x |g|^{1/2} K(x,x') \psi(x').
\eeq
It is symmetric in its arguments,
\beq
K(x,x') = K(x',x)
\eeq
as is evident from the fact that its complex conjugate $\ol{K}(x,x')$
(called the antipropagator) is the integral kernel of $(\wh C + i0)^{-1}$
and therefore equals also its adjoint. Equations (3.13) and (4.5) imply
\beq
(\eta \psi_0)(x) = \int d^4x |g|^{1/2} G_1(x,x') \ol{\psi}_0(x')
\eeq
where
\beq
G_1 := i(K - \ol{K})
\eeq
is a real symmetric ``Green function'' (solution of (2.4) in both 
arguments). Since $G_1$ is the kernel of the identity w.r.t.
$\langle \;,\;\rangle_{\rm phys}$,
\beq
\psi(x) = \langle G_1(x,\cdot),\psi\rangle_{\rm phys},
\eeq
it is the kernel of $N$ w.r.t. the charge form:
\beq
(N \psi)(x) = \langle G_1(x,\cdot),N \psi \rangle_{\rm phys} =
(G_1(x,\cdot),\psi)
\eeq
or
\beq
(N\psi)(x) = i \int d\sigma^a G_1(x,x') \st{\leftrightarrow}{\nabla}{}'_a
\psi(x) \equiv -i (G_1 * \psi)(x).
\eeq
For a further evaluation of (4.12) we make use of the chronological
decomposition of the Feynman propagator,
\beq
K(x,y) = \Theta(x,\Sigma(y))G^\uparrow(x,y) - \Theta(\Sigma(y),x)
G^\downarrow(x,y),
\eeq
where $\Sigma(y)$ is an arbitrary spacelike Cauchy hypersurface
containing $y$, the chronological step function $\Theta(x,\Sigma)$ is
one if $x$ is in the chronological future of $\Sigma$ and zero otherwise,
and $\Theta(\Sigma,x) = 1 - \Theta(x,\Sigma)$. The kernels $G^\uparrow$,
$G^\downarrow$ solve (2.4) in both arguments. If the domain $D$ of
$\wh C$ is characterized by asymptotic fall-off conditions (as is
generically the case), then the chronological decompositions of $K$ and
$\ol{K}$ define four projection operators:
\beqa
P^\uparrow \psi := G^\uparrow *\; \psi,&& \qquad \qquad
P^\downarrow \psi := G^\downarrow *\; \psi \\
\ol{P}^\uparrow \psi := \ol{G}^\uparrow *\; \psi,&& \qquad \qquad
\ol{P}^\downarrow \psi := \ol{G}^\downarrow *\; \psi .
\eeqa
They obey the relations
\beq
P = P^\dg
\eeq
\beq
P^\uparrow \; P^\downarrow = 0 = \ol{P}^\uparrow \; \ol{P}^\downarrow
\eeq
\beq
P^\uparrow + P^\downarrow = \mbox{id}_{H_{\rm phys}} =
\ol{P}^\uparrow + \ol{P}^\downarrow
\eeq
where the adjoint in (4.16) is defined with respect to
$\langle \;,\;\rangle_{\rm phys}$. For a derivation of these relations
as well as the projection property itself see \cite{R1}.

From (4.9) and (4.12) -- (4.15) we conclude that $N$ is the difference
of two projections
\beq
N = P^\uparrow - \ol{P}^\uparrow = \ol{P}^\downarrow - P^\downarrow,
\eeq
and hence (4.3) and the positivity condition (4.2) do indeed hold. The
stronger supremum condition (2.26) is valid exactly if the spectrum of
$N$ consists only of $+1$ and $-1$, i.e. if $\ol{P}^\downarrow =
P^\uparrow$ and $P^\downarrow = \ol{P}^\uparrow$. In this case the unique
Fock space obtained by setting $\langle\;,\;\rangle = 
\langle \;,\;\rangle_{\rm phys}$ in Sec.~2 is indeed physical, as has been
verified in concrete examples \cite{R3}. If (2.26) does not hold, a unique
irreducible Fock representation may still be defined by choosing $\Ha^+$
to consist of the eigenfunctions of $N$ corresponding to positive
eigenvalues (as was proposed by Nachtmann \cite{N}). This definition
does not even require the positivity condition (2.32). However it cannot
be correct in general, because strong physical arguments speak against
the existence of a unique natural ``vacuum'' state (e.g. it would
preclude the possibility of particle creation by the external field).

In general the physically correct procedure appears to be to use the
projection operators (4.14), (4.15) to define two Fock spaces
$\F_s^\uparrow$, $\F_s^\downarrow$ based on the one-particle Hilbert
spaces
\beq
\Ha^+ = P^\uparrow \; \Ha^+_{\rm phys}, \qquad
\Ha_+ = \ol{P}^\downarrow \; \Ha_{\rm phys}.
\eeq
Concrete examples \cite{R3} show that $\Ha_+$ and $\Ha^+$ define in-
and outgoing physical particles, respectively. In particular, particle
creation is predicted in this way for space-times where it is expected
for physical reasons. The Bogoliubov transformation between the in- and
outgoing representation is not unitarily implementable in general, and
neither the in- nor the out-vacuum will be Hadamard states (this being 
a local property in contrast to the global character of our definition
of states). Finally, although the mixed quasifree state $\omega$ associated 
with $\langle\;,\;\rangle_{\rm phys}$ via (2.33) is defined in general,
it does not seem to have a clear physical interpretation (it will
contain contributions of arbitrary particle number in $\F_s^\uparrow$
and $\F_s^\downarrow$).

\section{Concluding remarks: relativistic quantum mechanics}
As is well known, perturbative quantum field theory has an equivalent
formulation in relativistic quantum mechanics that was developed by
Feynman. The main ingredients for the calculation of amplitudes in this
approach are the charge form and the Feynman propagator (supplemented
by vertex rules in the interacting case). Remarkably, both of these
elements are dispensable once the inner product 
$\langle\;,\;\rangle_{\rm phys}$ is introduced. First of all, the 
one-particle Hilbert spaces may be defined without reference to the
Feynman propagator in the following way \cite{R1}: Solutions
$\psi^{(0)}$ in $\Ha^+$ ($\Ha_+$) are regular in the asymptotic future
(past) upon analytic continuation in the parameter $\lambda$ of the
eigenvalue equation $\wh C \psi^{(\lambda)} = \lambda \psi^{(\lambda)}$
into the upper (lower) half complex $\lambda$-plane. With this definition
of in- and outgoing particle (and antiparticle) states, it can be shown
\cite{R1} that the inner product $\langle\;,\;\rangle_{\rm phys}$ yields
directly the physical amplitudes. E.g. the relative amplitude for pair
creation in the mode ${}^+\psi$ is
$\langle{}^+\psi,{}^+\ol{\psi}\rangle/\langle{}^+\psi,{}^+\psi\rangle$.
This version of relativistic quantum mechanics is as self-contained as
the nonrelativistic theory (with e.g. pair creation treated as
``backscattering into the future''). Yet another approach to relativistic
quantum mechanics is the Hamiltonian path-integral quantization of the
relativistic particle \cite{H}. It is gratifying to observe that this
yields $\langle x|x'\rangle = K(x,x')$ where $K$ is the Feynman propagator
as defined in Sec.~4.

In conclusion, then, it appears that refined algebraic quantization does
indeed define a physical inner product and Hilbert space of solutions for
quantum field theory in curved space-time. The construction has to be
supplemented by a proper definition of states, which, however, is
suggested by the formalism itself. Moreover it shows that relativistic
quantum mechanics is more than just a prelude to quantum field theory.

\end{document}